\documentstyle[ecssm]{article}

\begin{document}

\title{Search for relativistic beaming in ultraluminous X--ray sources}

\author{L. Foschini\dag\footnote{E-mail: foschini@bo.iasf.cnr.it.}, 
G. Malaguti\dag, G. Di Cocco\dag, M. Cappi\dag, M. Dadina\dag, L.C. Ho\ddag}

\affil{\dag\ Istituto di Astrofisica Spaziale e Fisica Cosmica del CNR, Sezione di Bologna\\
(formerly iTeSRE--CNR). Via Gobetti 101 -- 40129 Bologna (Italy).}

\affil{\ddag\ The Observatories of the Carnegie Institution of Washington\\
813 Santa Barbara Street, Pasadena, CA 91101, USA}

\beginabstract
The nature of the ultraluminous X--ray sources (ULX) is still to be understood.
Several hypotheses have been suggested to explain the high X--ray luminosity:
stellar mass black holes with emission collimated by relativistic effects or
anisotropies in the accretion disk, intermediate mass black holes, 
young supernovae remnants, background active galactic nuclei. To discriminate
among the above hypotheses, the search for relativistic effects can be a 
valuable ``pathfinder''. 
Here we propose a list of ULX candidates suitable to show relativistic beaming.
\endabstract

\section{Introduction}
One of the most intriguing and exotic prediction of the theory of General
Relativity is the existence of a spacetime singularity with around it
a region bounded by an event horizon (i.e. a \emph{black hole}). Although the 
black hole was predicted early in 1916 by K. Schwarzschild (with some 
interesting pioneers such as Laplace), the proofs -- still indirect -- of 
its existence came only at the end of the XX Century, when the first 
satellites for X--ray astronomy were launched. 

Now, the black holes (BH) are divided into three categories, according to their mass:
stellar (e.g. Cyg X--1), intermediate, and supermassive (AGN). 
The existence of intermediate mass black holes was suggested after the discovery 
of the ultraluminous X--ray sources (ULX) and it is still controversial. 
The ULX are non--nuclear point--like sources with luminosity -- calculated as they
are in the host galaxy -- exceeding the Eddington limit for a neutron star (see
Foschini  \etal 2002a for a discussion on the luminosity threshold). They were known since
the \emph{Einstein} satellite (Fabbiano 1989), but it was with \emph{Chandra} 
that it was realized that ULX are common in nearby galaxies (Fabbiano \etal 2001). 

Now, ULX are a well known reality, but their nature is far from being known. 
Several hypotheses have been suggested: the most interesting would be that ULX are
intermediate mass black holes ($10^2-10^5M_{\odot}$, see e.g. Colbert and Mushotzky 1999, 
Miller \etal 2002). This could have a great impact also in the evolution of the host 
galaxies and provide a clue between the stellar mass BH and the supermassive BH in the AGN. 
Other hypotheses to explain ULX are: Kerr BH (Makishima \etal 2000), anisotropic 
emission from an accreting BH (King \etal 2001), slim accretion disk (Watarai \etal 2001), 
inhomogeneities in a radiation--pressure dominated accretion disk (Begelman 2002), emission from
a relativistic beamed jet (K\"ording \etal 2002, Georganopoulus \etal 2002), 
an extragalactic version of the microquasar model by Mirabel and Rodriguez (1999). 
There could be also the possibility that we are facing with young supernovae
remnants expanding in an interstellar matter rich of nitrogen (Schlegel 1994) or,
simply, with AGN of background (Foschini \etal 2002c).

Here we propose a list of ULX candidates suitable to display relativistic beaming, being
examples of microquasar in nearby galaxies. We would like to stress that the present list
is only a starting point for further researches, and therefore is far from being complete.

\section{What to search for and with what instrument}
From the observations of stellar mass black holes in the Milky Way, it is known that 
they can exist in different states. The two main are
named ``soft (high)'', where the thermal contribution from the accretion disk dominate
the X--ray emission, and ``hard (low)'', where the main contribution comes from 
a non--thermal power law (cf, for example, Nowak 1995). The hard state is generally
coupled with a relativistically beamed jet with optically thick emission of synchrotron
radiation in the radio--optical domain (Fender 2001), but it has been recently proposed that 
this radiation can extend up to the X--ray domain. In this case, the emitted radiation can be generated
either in the extended jet structure (Atoyan and Aharonian 1999) or in the region close to
the accretion disk (Markoff \etal 2001). Other mechanism able to produce X--ray photons
are the inverse Compton scattering of the relativistic electrons from external seed 
photons (from the companion star, e.g. Georganopoulos \etal 2002) or from the photons of the
accretion disk (cf Nowak \etal 2002). In the observation of ULX we move to nearby galaxies, 
so that it is much more difficult to find the proofs of a relativistically beamed emission
with the actual technology. Nonetheless, it is possible to set up a first tentative scenario.

\subsection{Radio emission}
The easiest way could be to search for the radio emission, but when scaled to the distances
of the nearby galaxies, we realize that it is a task not so easy (and indeed, the radio emission
from ULX is still poorly known).

One of the most known Galactic microquasar candidates, GRS$1915+105$, is characterized by episodes 
of radio flares up to $1$~Jy during its active state, which are superimposed to a ``plateau'' of about
$100$~mJy at $\nu \approx 1-10$~GHz (see Atoyan and Aharonian 1999). Scaling the distance to
a typical neaby galaxy $5$~Mpc far, the peak flux drops down to $6$~$\mu$Jy (and this would be the best case). 

There are actually a very few cases of radio emission from ULX: the first case was later
identified as a BL Lac of background (Foschini \etal 2002c). Another case has been 
recently found by Kaaret \etal (2003): the ultraluminous source 2E~$1400.2-4018$ in NGC5408 
showed a flux of $0.26$~mJy at $4.8$~GHz and the authors suggested that this could be -- together
with the X--ray spectrum (from Chandra data) -- the signature of a microquasar. 
If confirmed (no redshift was available at the date of the publication), this would 
represent a very powerful ULX, about two order of magnitude higher than the peak of 
GRS~$1915+105$ when scaled to the same distance. The most promising radio
detection appears to be that of ULX M81--X6: $95$~$\mu$Jy at $\lambda = 3.6$~cm 
(Swartz \etal 2003). So, the hunt for the radio emission is still well open.

\subsection{X--ray emission}
Most of ULX show thermal emission well fitted with the multicolor blackbody accretion disk
(MCD, cf Makishima \etal 2000). As known, the thermal contribution in the hard state 
is generally below the 20\% of the total emission in the X--ray domain; in addition, 
the MCD model can also be explained well by the emission from the innermost stable 
circular orbit of the accretion disk of an intermediate mass BH (Miller \etal 2002) 
or from a spinning BH (e.g. La Parola \etal 2001), depending on if the disk is cold 
($kT\approx 0.15$~keV) or hot ($kT\approx 1-2$~keV), respectively.

Therefore, to search for the evidence of jets, the best option appears to be to look at 
those sources displaying spectra fitted with power law models. In addition, 
the presence of relativistic beaming could be inferred by the absence of 
reflection features.

Moreover, it is worth noting that the hard state displays an evident time variability 
(up to 50\% rms below 1~Hz, cf Fender 2001). However, this topic has been 
investigated poorly, because of the necessity to have very long exposure, 
possibly spaced by months or years (e.g. Fabbiano \etal 2003).

\subsection{Caveats: how different detectors see the spectrum of the sources}
The spectral data available on ULX are mainly from three satellites: ASCA, Chandra, and 
XMM--Newton. ASCA had high throughput of photons, but the low spatial resolution make
it possible to have often contamination from other nearby sources. Chandra, while 
having an exceptional angular resolution ($< 1''$), has an effective area that drops above 
the peak energy of the multicolor disk blackbody model generally used for ULX 
($kT\approx 1-2$~keV); in addition, with comparison to ASCA effective area, there is 
another sharper drop in the effective area just above $4$~keV. On the other hand,
XMM--Newton appears to be a good compromise between spatial and spectral behaviour to observe
and study the jetted ULX in nearby galaxies: 
while the angular resolution of the EPIC--MOS cameras is around $1.1''$, the effective
areas -- particularly of the PN camera -- can guarantee a high throughput also at energies
above $2$~keV, so providing good measurement in the region of hard--X rays, where it is 
expected to be dominated by the tail of the power law.

\section{A list of candidates}
As stated in the Sect.~2.3, ASCA has a very low spatial resolution, although the high
throughput allows to get good spectra (e.g. Makishima \etal 2000, Mizuno \etal 2001). 
Sometimes ULX are confused with a low luminosity AGN or, alternatively, 
that the observed ``low luminosity AGN'' is, in reality, an ULX. Two cases of this type 
occurred in NGC4565 and NGC3486 respectively (cf Foschini \etal 2002b). So, to avoid
further contaminations, we adopted the -- perhaps too much -- drastic decision of not 
using the ASCA data. We use, as starting point, the Chandra and XMM--Newton data, but
it is worth noting that ASCA observations could be a good ``pathfinder'' for further researches.

In the Table~1, a list of ULX candidates with possible relativistic beaming as inferred
from the power law model. In the hard state, the photon index is $1.3-1.9$ (cf Ebisawa \etal 1996, 
Fender 2001) and we considered as possible candidates those sources with $\Gamma$ in the above
interval, also by taking into account the statistical errors. It is worth noting that Zezas \etal (2002)
have found in the Antennae galaxies several other ULX best fitted with a power law model, but the
$\Gamma$ values were incredibly high (up to 7!). We do not consider these sources, but given 
the characteristics of the effective area of Chandra, it is possible to guess that the same sources
when observed with XMM--Newton could display flatter indexes. Therefore, these sources (namely, n. 5, 13,
29, 33) need for additional investigation.

\begin{table} 
\footnotesize\rm 
\caption{List of ULX candidates with possible relativistic beaming as inferred
from spectral characteristics (i.e., best fit with a power law model). Columns: 
(1) Name of the source; (2) right ascension and declination (J2000); 
(3) absorbing column density [$10^{21}$ cm$^{-2}$]; (4) photon index; (5) 
X--ray luminosity in the $0.5-10$ keV energy band [$10^{38}$ erg s$^{-1}$]; (6) reference paper; 
(7) X--ray satellite that made the observation.  
For $N_{\mathrm{H}}=$Gal., it means that no additional absorption is found to be significant.
The luminosities were calculated using distances from Ho \etal (1997), 
who considered an infall velocity of 300 km/s for the Local Group,
$H_0=75$~km$\cdot$s$^{-1}$Mpc$^{-1}$, and the distance of the Virgo Cluster 
of 16.8~Mpc.}  
\label{powerlaw} 
\begin{center}        
\begin{tabular}{lllllll}
\topline 
Name & Coordinates & $N_H$ & $\Gamma$ & $L$ & Reference & Satellite\\
(1)  & (2)         & (3)   & (4)      & (5) & (6)       & (7) \\  
\midline
Antennae--37  & $12:01:55.0,-18:53:15$ & $2.6\pm0.5$ & $1.7\pm0.2$ & $102$ & Zezas \etal 2002 & Chandra\\
Antennae--44  & $12:01:56.4,-18:51:58$ & $1.0\pm0.3$ & $1.9\pm0.1$ & $129$ & Zezas \etal 2002 & Chandra\\
NGC720--40    & $01:53:01.7,-13:46:31$ & $<4$     & $1.3\pm 0.7$ & $110$ & Jeltema \etal (2002) & Chandra\\ 
NGC3486--ULX1 & $11:00:22.4,+28:58:18$ & Gal.     & $2.2\pm 0.5$ & $5.0$ & Foschini \etal (2002b) & XMM\\
NGC3941--ULX1 & $11:52:58.3,+36:59:00$ & Gal.     & $1.9\pm 0.2$ & $74$ & Foschini \etal (2002b) & XMM\\
NGC4501--ULX2 & $12:32:00.8,+14:24:42$ & Gal.     & $2.3\pm 0.4$ & $37$ & Foschini \etal (2002b) & XMM\\
NGC4565--ULX2 & $12:36:14.8,+26:00:53$ & $6\pm 5$ & $1.7\pm 0.6$ & $16$ & Foschini \etal (2002b) & XMM\\
NGC4565--ULX4 & $12:36:17.4,+25:58:54$ & Gal.     & $1.9\pm 0.1$ & $25$ & Foschini \etal (2002b) & XMM\\
NGC4565--ULX6 & $12:36:27.8,+25:57:34$ & Gal.     & $1.5\pm 0.3$ & $9.0$ & Foschini \etal (2002b) & XMM\\
J123030.8+413911 & $12:30:30.8,+41:39:11$ & $8\pm3$ & $1.8\pm 0.3$ & $46$ & Roberts \etal 2002 & Chandra\\
(NGC4485/4490)   & {} & {} & {} & {} & {} & {}\\
NGC4490--X1 & $12:30:36.3,+41:38:37$ & $4\pm2$ & $1.8\pm 0.3$ & $26$ & Roberts \etal 2002 & Chandra\\
NGC4490--X2 & $12:30:32.8,+41:39:18$ & $5\pm2$ & $1.7\pm 0.3$ & $26$ & Roberts \etal 2002 & Chandra\\
   
\bottomline 
\end{tabular} 
\end{center} 
\end{table} 

In addition to the ULX of Table~1, we also propose another set of candidates to display relativistic
beaming, but now as inferred from time behaviour. In this case, the best fit model is no more
the power law, but we also consider the blackbody models. The presence of the relativistic
beaming is now inferred from the high time variability (up to 50\% rms). 
This second set is shown in Table~2.

\begin{table} 
\footnotesize\rm 
\caption{List of ULX candidates with possible relativistic beaming as inferred
from time variability studies. Columns: 
(1) Name of the source; (2) right ascension and declination (J2000); 
(3) absorbing column density [$10^{21}$ cm$^{-2}$]; (4) photon index, if the best fit is a power law,
or $kT$ [keV] in the case of blackbody; (5) 
X--ray luminosity in the $0.3-8$ keV energy band [$10^{38}$ erg s$^{-1}$]; (6) reference paper; 
(7) X--ray satellite that made the observation.  
For $N_{\mathrm{H}}=$Gal., it means that no additional absorption is found to be significant.
The luminosities were calculated using distances from Ho \etal (1997), 
who considered an infall velocity of 300 km/s for the Local Group,
$H_0=75$~km$\cdot$s$^{-1}$Mpc$^{-1}$, and the distance of the Virgo Cluster 
of 16.8~Mpc.}  
\label{timing} 
\begin{center}        
\begin{tabular}{lllllll}
\topline 
Name & Coordinates & $N_H$ & $\Gamma$ or $kT$ & $L$ & Reference & Satellite\\
(1)  & (2)         & (3)   & (4)      & (5) & (6)       & (7) \\  
\midline 
2E$1400.2-4108$ & $14:03:19.6,-41:22:59$ & $4\pm2$ & $\Gamma=2.8\pm0.3$ & $110$ & Kaaret \etal (2003) & Chandra\\
(NGC5408)  & {} & {} & {} & {} & {} & {}\\
M51--X7       & $13:30:01.0,+47:13:44$ & Gal.         & $\Gamma=1.3\pm 0.2$ & $19.8$ & Liu \etal (2002) & Chandra\\
M81--X6       & $09:55:33.0,+69:00:33$ & $2.2\pm 0.1$ & $kT=1.0\pm 0.1$ & $27$ & Swartz \etal (2003) & Chandra\\
M101--P104    & $14:03:36.1,+54:19:25$ & Gal.         & $\Gamma=2.6$    & $2.0$ & Mukai \etal (2003) & Chandra\\    
\bottomline 
\end{tabular} 
\end{center} 
\end{table} 

\section{Final remarks}
Even though during the latest years, the efforts to understand the nature of ULX dramatically increased,
these enigmatic sources are still poorly known. Among the different available theories, it is worth 
mentioning the microquasar model. To search for proofs in favour of this model, it is necessary
to study the spectral and timing behaviour of the ULX with dedicated observations. The list of
ULX presented in this work is a first step toward this direction.

\section*{Acknowledgments}
This research has made use of NASA's Astrophysics Data System.
LF acknowledge G. Ghisellini for useful discussions.

\end{document}